\documentclass[nofootinbib, showkeys, twocolumn, a4paper, 10pt, aps, prd, preprintnumbers, superscriptaddress, floatfix]{revtex4-2}
\pdfoutput=1
\usepackage{amsmath,amssymb,graphicx,bm,psfrag,color,slashed,array,subfiles,cancel,braket,multirow,bigstrut}
\usepackage[version=4]{mhchem}
\usepackage[dvipsnames]{xcolor}
\usepackage[font=footnotesize,labelfont=bf,skip=5pt,justification=RaggedRight]{caption}
\usepackage{subcaption}
\usepackage{hyperref}
\hypersetup{
    colorlinks=true,
    linkcolor=blue,
    filecolor=magenta,      
    urlcolor=cyan,
    pdftitle={Projected sensitivity of the ANUBIS detector to Heavy Neutral Leptons},
    pdfpagemode=FullScreen,
    }
\usepackage{xspace} 
\usepackage{verbatim} 
\usepackage{siunitx} 
\DeclareSIUnit{\year}{y}

\usepackage{enumitem}
\setlist{itemsep=0mm}

\newcommand{\eg}{\mbox{\itshape e.g.}\xspace}
\newcommand{\ie}{\mbox{\itshape i.e.}\xspace}

\newcommand{\anubis}{\mbox{ANUBIS}\xspace}
\newcommand{\setanubis}{\mbox{\textsc{SET-ANUBIS}}\xspace}

\newcommand{\atlas}{\mbox{ATLAS}\xspace}

\newcommand{\codex}{\mbox{CODEX-b}\xspace}

\newcommand{\pythia}[1][]{\mbox{\textsc{Pythia#1}}\xspace}
\newcommand{\madgraph}{\mbox{\textsc{MadGraph}}\xspace}
\newcommand{\madspin}{\mbox{\textsc{MadSpin}}\xspace}
\newcommand{\madversion}[1]{MG5\_aMC@NLO #1}

\newcommand{\mathematica}{\mbox{\textsc{Mathematica}}\xspace}

\newcommand{\geant}{\mbox{\textsc{Geant4}}\xspace}

\newcommand{\mm}{\ensuremath{\textnormal{mm}}\xspace}

\newcommand{\metre}{\ensuremath{\textnormal{m}}\xspace}

\newcommand{\GeV}{\ensuremath{\textnormal{GeV}}\xspace}

\newcommand{\met}{\ensuremath{E_T^\text{miss}}\xspace} % Missing Transverse Energy
\newcommand{\pt}{\ensuremath{p_T}\xspace} % Transverse momentum
\newcommand{\order}[1]{\ensuremath{\mathcal{O}(#1)}\xspace} % Order of X
\newcommand{\acc}{\ensuremath{\mathcal A}\xspace} %Acceptance
\let\BF=\br
\def\pmbanner{}
\begin{document}
\title{\pmbanner{\LARGE \texorpdfstring{Projected sensitivity of the \anubis detector to Heavy Neutral Leptons\\ 
\includegraphics[width=0.1\linewidth]{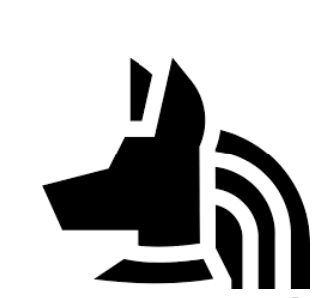}\\
\vspace{-3mm}
{\large\textbf{ANUBIS Collaboration}}
\\[-0.2em] \vspace{-1mm}
{\normalsize\textit{E-mail: }\href{mailto:anubis-publications@cern.ch}{anubis-publications@cern.ch}}
\\[-0.2em]
}{Projected sensitivity of the ANUBIS detector to heavy neutral leptons}}
}

% \collaboration{\textbf{ANUBIS Collaboration}}
% \email{anubis-publications@cern.ch}

\begin{abstract}
%    \linenumbers
    \textit{Abstract:} 
    Long-Lived Particles (LLPs) are a common feature in various extensions to the Standard Model (SM) that seek to address known limitations. 
    The \anubis detector has been proposed to extend the sensitivity of the ATLAS experiment at the LHC to LLPs by instrumenting the ceiling of the ATLAS detector cavern.
    This article presents the projected sensitivity of \anubis to Heavy Neutral Leptons (HNLs). 
    For a minimal Majorana HNL model that only couples to a single flavour of lepton ($e$ or $\mu$) \anubis reaches a maximum sensitivity of $|V_{1e}|^2=1.8\times10^{-8}$ and $|V_{1\mu}|^2=1.9\times10^{-8}$ for a HNL mass of $m_{N_1}=6.4~\GeV$ and 6.3~\GeV respectively. This provides complementary coverage to other proposed LLP experiments in the HNL parameter-space, with potential for significant improvement during ANUBIS data-taking through advances in analysis strategies.
    The results are obtained with \setanubis, a flexible framework to evaluate the sensitivity of \anubis to a variety of LLP models.
%    \begin{center}
%        Version 3.0\\[-0.3em]
%        \today 
%    \end{center}
\end{abstract}

%\date{}
% authors (by surname)
\author{Martin Bauer}
\affiliation{IPPP, Dept of Physics, Durham University, Durham, United Kingdom}%
\author{Rachel Bentham}
\affiliation{Department of Physics and Astronomy, University of Manchester, Manchester, United Kingdom}%
\affiliation{Formerly at: Cavendish Laboratory, University of Cambridge, Cambridge, United Kingdom}
\author{Oleg Brandt}
\affiliation{Cavendish Laboratory, University of Cambridge, Cambridge, United Kingdom}
\author{Sofie Nordahl Erner}
\affiliation{Formerly at: IPPP, Dept of Physics, Durham University, Durham, United Kingdom}
\author{Anna Mullin}
\affiliation{Formerly at: Cavendish Laboratory, University of Cambridge, Cambridge, United Kingdom}
\author{David Peng}
\affiliation{Formerly at: Cavendish Laboratory, University of Cambridge, Cambridge, United Kingdom}
\author{Th\'eo Reymermier}
\affiliation{Formerly at: Cavendish Laboratory, University of Cambridge, Cambridge, United Kingdom}
\affiliation{Institut de Physique des 2 Infinis (IP2I), Lyon, France}
\author{and Paul Swallow}
\affiliation{Cavendish Laboratory, University of Cambridge, Cambridge, United Kingdom}
\author{for the ANUBIS Collaboration}
\noaffiliation

\def\andname{\hspace*{-0.5em}}
\maketitle

%%%%%%%%%%%%%%%%%%%%%%%%%%%%%%%%%%%%%%%
\section{Introduction} \label{sec:intro}
%%%%%%%%%%%%%%%%%%%%%%%%%%%%%%%%%%%%%%%

Among the various experimental searches for new particles and new interactions, the study of Long-Lived Particles (LLPs) presents a compelling prospect. 
Such particles are characterised by proper lifetimes that are long ($\tau>10$~ps) compared to most Standard Model (SM) particles and arise generically in many Beyond the Standard Model (BSM) theories. 
LLPs are predicted in a multitude of BSM scenarios~\cite{Antel:2023hkf}, and can provide insights into Dark Matter (DM)~\cite{Hall:2010jx,Cheung:2010gk,Zurek:2013wia}, non-zero neutrino masses~\cite{Bondarenko:2018ptm}, the strong CP problem~\cite{PhysRevLett.38.1440}, and other phenomena, \eg neutral naturalness~\cite{Giudice:1998bp,Burdman:2006tz}. 
However, the macroscopic lifetime and other LLP properties such as weak couplings to SM matter pose major experimental challenges that necessitate innovative detection strategies and/or dedicated detector designs.

The \anubis detector foresees searching for LLPs by instrumenting the ceiling of the \atlas UX15 experimental cavern with tracking detectors~\cite{Bauer:2019vqk,ANUBIS:2025sgg}. 
The \anubis detector would work symbiotically with the \atlas experiment~\cite{Aad:1129811} to extend its sensitivity to LLPs with lifetimes ranging from $c\tau\approx\order{1}~\metre$ up to $\mathcal{O}(10^5)~\metre$ by increasing the geometrical acceptance to displaced LLP decays~\cite{ATLAS:2013dyx}. 

As a transverse detector, \ie located transverse to the beamline, \anubis targets neutral LLPs that are produced at partonic centre of mass energies, $\sqrt{\hat{s}}$, at the electroweak scale and above. 
It also targets long LLP lifetimes by exploiting its large geometrical acceptance. This makes \anubis complementary to general-purpose detectors like \atlas~\cite{Aad:1129811} or CMS~\cite{Chatrchyan:1129810}, as well as forward detectors such as FASER~\cite{FASER:2018eoc}, or beam-dump experiments like SHiP~\cite{Alekhin:2015byh}. 
%which are more sensitive to lighter LLPs and cannot reach the electroweak scale in $\sqrt{\hat{s}}$. While several forward LLP detectors are currently operational~\cite{FASER:2018eoc}, 
Currently, no transverse LLP detectors exist, highlighting the unique physics potential of \anubis, alongside other proposals like \codex~\cite{Gligorov:2017nwh,Aielli:2019ivi} and MATHUSLA~\cite{MATHUSLA:2019qpy,MATHUSLA:2025zyt}.
%There are several proposals including \anubis, such as CODEX-b~\cite{Gligorov:2017nwh,Aielli:2019ivi} and MATHUSLA~\cite{MATHUSLA:2019qpy,MATHUSLA:2025zyt}, one of which should be constructed to allow for full exploitation of the LLP parameter-space.

\begin{figure*}[htb]
    \begin{center}
    \begin{subfigure}[b]{0.49\linewidth}
        \centering
        \includegraphics[height=0.67\linewidth]{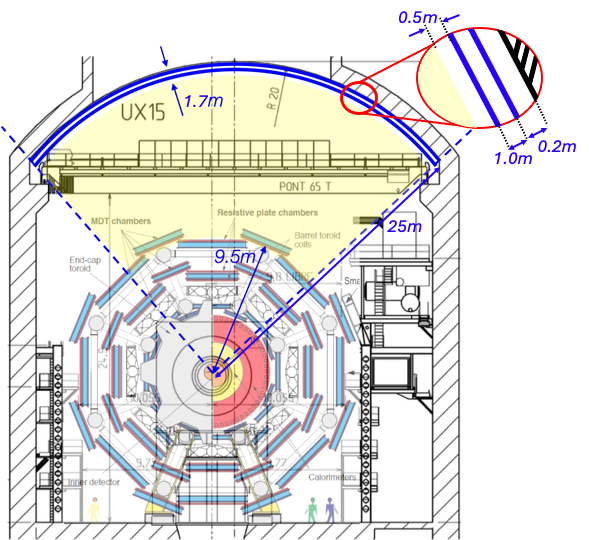}
        \caption{}
        \label{fig:ANUBISAcceptance_XY}
    \end{subfigure}
    \begin{subfigure}[b]{0.49\linewidth}
        \centering
        \includegraphics[height=0.7\linewidth]{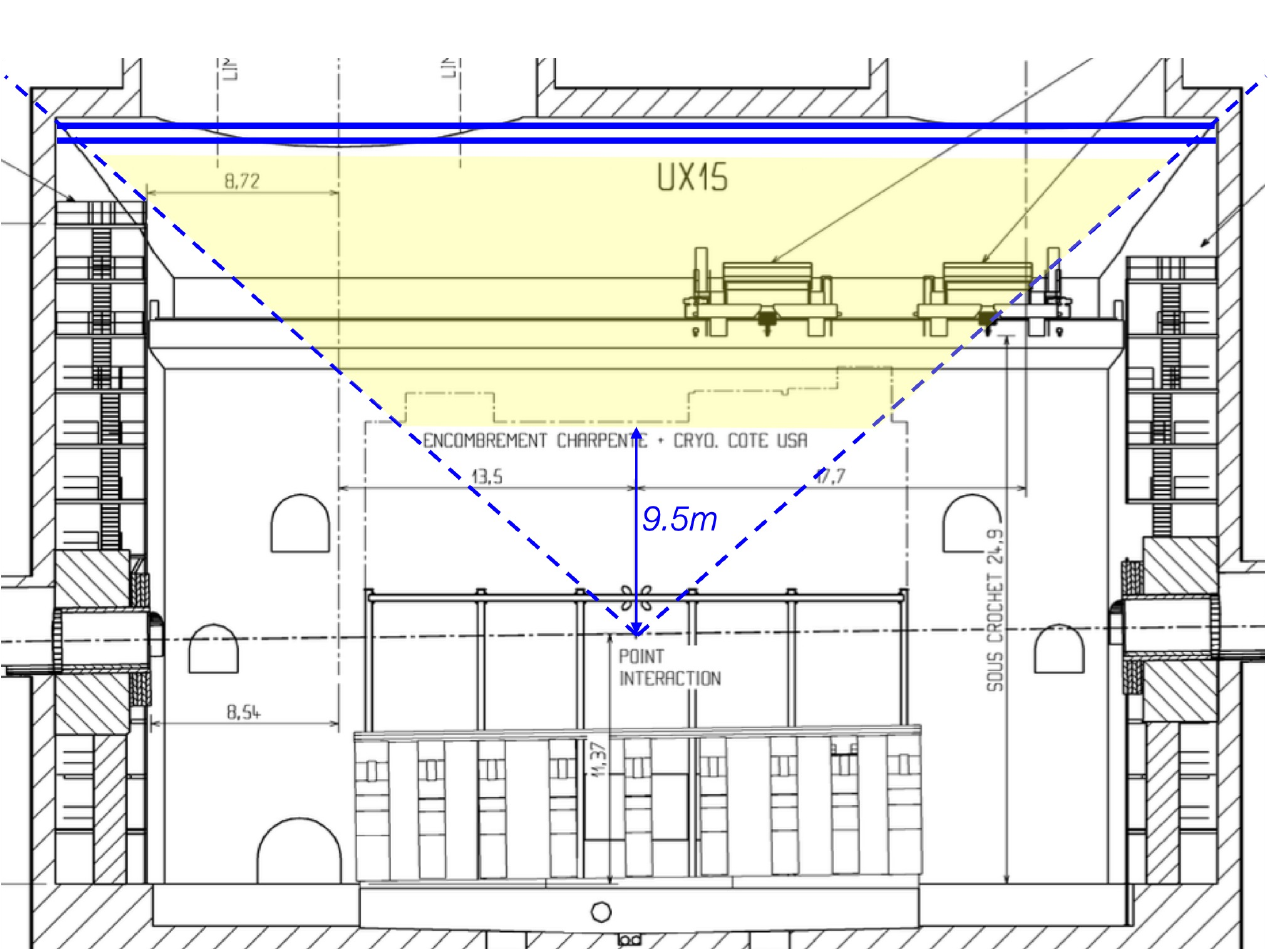}
        \caption{}
        \label{fig:ANUBISAcceptance_ZY}
    \end{subfigure}
    \end{center}
    \caption{
    The two layers of the tracking stations of the \anubis detector for the ceiling configuration are shown as blue thick lines inside the UX15 ATLAS experimental cavern alongside the \atlas detector in (a)~the $(x,y)$ plane and (b)~the $(y,z)$ plane. 
    The fiducial volume is indicated as a shaded yellow region and the acceptance boundaries in $\eta$ and $\phi$ are highlighted as dashed blue lines. 
    The drawings of the \atlas detector and the cavern are courtesy of \atlas.}
    \label{fig:ANUBISAcceptance}
\end{figure*}

The \anubis tracking detector system is composed of a set of two curved layers separated by 1~\metre, where each layer contains modules with a triplet of Resistive Plate Chambers (RPCs).
An additional singlet layer is being studied to improve pattern recognition, but is not considered in this article. 
The radial extent of the fiducial volume of \anubis to reconstruct displaced vertices from LLP decays is defined from 9.5~m away from the ATLAS beam line to about 23~m, \ie 1.7~m below the ceiling and hence 0.3~m below the lowest tracking layer~\cite{ANUBIS:2025sgg}, as shown in Figure~\ref{fig:ANUBISAcceptance}. 

The \anubis detector is planned to be fully integrated with the \atlas trigger system so that when \anubis registers an LLP candidate event it can trigger the \atlas detector, and vice versa. 
The full view of the event obtained with the ATLAS and ANUBIS detectors will give valuable insights into the LLP production process and can provide options for additional selections for data analysis. 

Several types of background sources could contribute to LLP searches with the \anubis detector: cosmic rays; beam-induced backgrounds; decays of quasi-thermal neutrons; random coincidences of unrelated tracks; and neutral SM particles with macroscopic lifetimes like $K_L$ and $n$~\cite{ANUBIS:2025sgg}. 
The first three sources can be effectively rejected using the precise timing resolution of \anubis' tracking layers of 350~ps~\cite{ANUBIS:2026jxg} and their event topology. 
%For example, cosmic rays can be distinguished as down-going tracks and thus vetoed. 
The dominant background source is from $K_L$ and $n$, which can decay or undergo material interactions to produce hadrons. 
This background can be reduced through passive shielding of \atlas, as well as by exploiting the predominantly air-filled decay volume that strongly reduces the likelihood of hadronic interactions. 
Additionally, \atlas acts as an active veto to reject charged SM LLPs using kinematic and topological requirements described in Section~\ref{sec:SETacceptance}. 
Backgrounds have been conservatively determined to contribute at most 36 events for an exclusion sensitivity at 95\% confidence level (CL) using a data-driven background estimate~\cite{ANUBIS:2025sgg}. 
%Although this may be reduced through exploiting particular decay topologies, \eg with an associated prompt object that \atlas can trigger on.

This article is structured as follows. 
Section~\ref{sec:theo} provides an overview of the minimal Majorana Heavy Neutral Lepton (HNL) model studied in this article; the charge conjugate states for production and decay modes are also considered throughout. 
This is followed by an overview of the \setanubis framework and how it can be used to determine sensitivity limits for the \anubis detector in Section~\ref{sec:SETANUBIS}. Finally, Sections~\ref{sec:EventGen} and~\ref{sec:sens} detail the work performed to study the HNL model and determine the sensitivity limits of the \anubis detector to this model, respectively. 

%%%%%%%%%%%%%%%%%%%%%%%%%%%%%%%%%%%%%%%
\section{Heavy Neutral Leptons} 
\label{sec:theo}
%%%%%%%%%%%%%%%%%%%%%%%%%%%%%%%%%%%%%%%

To facilitate consistent comparisons between different experimental proposals, the Physics Beyond Collider (PBC) group has defined a set of Benchmark Cases (BC) that cover the four main types of LLP portals: via vector or scalar/Higgs mediators, neutrinos, and axions~\cite{Beacham:2019nyx,Alemany:2019vsk}. 
Of these, \anubis is expected to have good sensitivity to scenarios involving heavy mediators like the scalar and neutrino portals, \eg exotic Higgs sectors and Heavy Neutral Leptons respectively. 

This study focuses on the HNL portal to a hidden sector. 
Such models are of particular interest due to their minimal extension of the SM that can give rise to massive neutrinos by introducing $\mathcal{N}$ additional right-handed, neutral leptons ($N_i$ with $i \in [1,\mathcal{N}]$) using a seesaw model~\cite{Brdar_2019}. 
The minimal interactions between the HNL and SM particles arise through mixing with the active SM neutrinos. 
Such couplings can result in a macroscopic lifetime of the HNL, making it a prime LLP candidate. 
Due to this and the neutral electric charge of the HNLs, large regions of HNL parameter-space remain unexplored~\cite{Antel:2023hkf}.

Experimentally, neutrinos have been measured to be massive~\cite{Super-Kamiokande:1998kpq}, which is not predicted by the SM. 
In order to introduce neutrino masses into the SM while respecting all gauge invariances, a new field can be added that is a singlet of all SM gauge groups. This gives an additional contribution, $\mathcal{L}_{\text{HNL}}$, to the SM Lagrangian, $\mathcal{L}_{\text{SM}}$,
\begin{equation}
    \mathcal{L} = \mathcal{L}_{\text{SM}} + \mathcal{L}_{\text{HNL}},
\end{equation}
with
\begin{equation}
    \mathcal{L}_{\text{HNL}} = \frac{i}{2}\overline{N_i} \slashed{\partial}N_i -\frac{m_{N_i}}{2} \overline{N_i} N_i^c-y_{i \alpha} \overline{N_i} \tilde{\phi}^\dagger L^\alpha + h.c.,
    \label{eq:HNL_Lagrangian}
\end{equation}
where $\alpha \in \{e,\mu,\tau\}$, $N_i$ is the HNL field with the mass $m_{Ni}$, and $y_{i\alpha}$ is a matrix of complex Yukawa couplings~\cite{Li_2023,Brdar_2019}. 
Here, $L^\alpha$ is the SM left-handed leptonic doublet field, $\phi$ is the Higgs doublet field with $\tilde{\phi} = i\sigma_2 \phi^{*}$ for the second Pauli matrix $\sigma_2$, and $N_i^c = \gamma^2 N_i^{*}$ with $\gamma^2$ the second Dirac matrix. As the new field is a singlet under all SM gauge groups and has no associated charge, the covariant derivative for the new field simplifies to the derivative, $\slashed{\partial}$. A Majorana mass term for the field is introduced, and interactions with the SM particles arise through Yukawa interactions via mixing with the SM neutrino. 
The Dirac HNL scenario, obtained by omitting the Majorana mass term in Equation~\eqref{eq:HNL_Lagrangian}, is not explicitly considered in this work, but is expected to be qualitatively similar to the Majorana HNL case up to an $\mathcal{O}(1)$ factor from the decay width.

In this article, a single HNL that mixes with only one flavour of SM neutrino is introduced, \ie $i=1$ and ${\alpha=e,\mu}$ or $\tau$. 
This can be easily extended to a larger number of HNLs due to the new field being a gauge singlet.
%The Higgs field acquires a non-zero Vacuum expectation Value $v$.
%$v = \sqrt{-\mu^2 / \lambda}$ where $\mu$ and $\lambda$ are real parameters from the Higgs potential\footnote{Explicitly: $V = \mu^2 |\phi|^2 +\lambda |\phi|^4$ with $\phi = \phi_1+i\phi_2$.} and $\lambda{\text{,}}\,\mu < 0$. 
%
Choosing ${\phi_0 = \frac{1}{\sqrt{2}} \begin{pmatrix} 0 \\ v \end{pmatrix}}$ for the minimum after electroweak symmetry breaking, where $v$ is the vacuum expectation value of the Higgs field, the terms that contribute to the neutrino mass can be written as
%
% \begin{equation}
%     \mathcal{L}_{\text{HNL}}^{mass} \supset \frac{i}{2}\overline{N}_1 \slashed{D}N_1 -\frac{m_{N_1}}{2}\overline{N}_1^c N_1 -C_{1j} \overline{N}^1 \frac{v}{\sqrt{2}} L^j + \text{h.c}\,.
%     \label{eq:HNL_Lagrangian_VeV}
% \end{equation}
\begin{equation}
    \mathcal{L}_{\text{HNL}}^{M} \supset -\frac{y_{1\alpha} v}{\sqrt{2}} \overline{N_1}\nu_\alpha^L  -\frac{m_{N_1}}{2}\overline{N_1} N_1^c + h.c.,
    \label{eq:HNL_Lagrangian_VeV}
\end{equation}
where $\nu_\alpha^L$ is the component of the SM left-handed lepton doublet corresponding to the neutrino. 
Hence, the $\frac{y_{1\alpha} v}{\sqrt{2}}$ term can be identified as the Dirac mass, $m_D$, generated by electroweak symmetry breaking.
The Lagrangian in Equation~\eqref{eq:HNL_Lagrangian_VeV} can then be expressed in the flavour basis, $(\nu_\alpha^L, N_1^c)$, as~\cite{Brdar_2019}
\begin{equation}
    \mathcal{L}_{\text{HNL}}^{M} \supset -\frac{1}{2}\begin{pmatrix}\nu_\alpha^L & N_1^c\end{pmatrix}\begin{pmatrix} 0 & \frac{y_{1\alpha} v}{\sqrt{2}} \\ \frac{y_{1\alpha}v}{\sqrt{2}} & m_{N_1} \end{pmatrix}\begin{pmatrix}
        \overline{\nu_\alpha^L}\\ \overline{N_1}
    \end{pmatrix} + h.c.\,,
\end{equation}
where the mass matrix is identified as
\begin{equation}
    M_{\alpha1} = \begin{pmatrix} 0 & \frac{y_{1\alpha} v}{\sqrt{2}} \\ \frac{y_{1\alpha}v}{\sqrt{2}} & m_{N_1} \end{pmatrix} \,.
\end{equation}
After diagonalisation of the mass matrix, the neutrino mass eigenstates are obtained as 
\begin{equation}
    \lambda_\pm = \frac{m_{N_1}}{2}\left[1\pm\sqrt{1+2\left(\frac{|y_{1\alpha}|v}{m_{N_1}}\right)^2}\right],
\end{equation}
which can be associated with the mass eigenvalues of the SM neutrino\footnote{The $m_{\nu_\alpha^L}$ eigenvalue can be made positive by applying a rotation $\phi\rightarrow i\phi$.}, $m_{\nu_\alpha^L}$, and the HNL via a type-I seesaw mechanism that assumes $m_{N_1} \gg \frac{|y_{1\alpha}|v}{\sqrt{2}}$~\cite{Brdar_2019}. In that case, one obtains
\begin{align}
&m_{\nu_\alpha^L} 
%~\frac{m_{N_1} + \sqrt{m_{N_1}^2-4\left(\frac{y_{1\alpha} v}{\sqrt{2}}\right)^2}}{2} 
\approx -\frac{\left(\frac{|y_{1\alpha}|v}{\sqrt{2}}\right)^2}{m_{N_1}}\,,
%&m_{N_1} 
%= \frac{m_{N_1} + \sqrt{m_{N_1}^2+4\left(\frac{y_{1\alpha} v}{\sqrt{2}}\right)^2}}{2} 
%\approx m_{N_1}\,,
\label{eq:massEigenvalues}
\end{align}
where $m_{\nu_\alpha^L}$ naturally becomes very light if $m_{N_1}$ is large. 
%In the gauge eigenstate basis, HNLs do not mix with other SM particles through Z and W boson, they only mix through Higgs coupling. 

HNLs are coupled to SM neutrinos in the mass eigenstate basis through mixing
%with the relations,
%
% \begin{align}
% &\nu^m_l = \cos{(\theta_{1\alpha})} \, \nu^L_{\alpha} +\sin{(\theta_{1\alpha})} \, N_1^c, \\
% &N^m_l = \cos{(\theta_{1\alpha})} \, N_1 -\sin{(\theta_{1\alpha})} \, (\nu^L_{\alpha})^c, 
% \end{align}
% %
% where 
with the associated mixing parameter~\cite{Li_2023} by
\begin{align}
    |V_{1\alpha}|^2&=\sin^2{\theta_{1\alpha}} = \left(\frac{|y_{1\alpha}|v}{\sqrt{2}\,m_{N_1}}\right)^2.
\end{align}
This naturally results in a small coupling of the HNL, which in turn leads to it becoming a long-lived particle.
Typically, $|V_{1\alpha}|^2$ and $m_{N_1}$ are free parameters in HNL seesaw models.

The interactions of the HNL with the $W$ and $Z$ bosons arise from the mixing with the SM leptons via
\begin{align}
\label{eq:cc}
\mathcal{L}_{\text{HNL}}^W &= \frac{g V_{1\alpha}}{\sqrt{2}} \left[ W_\mu(\overline{\ell}_{1\alpha}^L \gamma^\mu N_1)\right]+h.c., \\ 
\label{eq:nc}
\mathcal{L}_{\text{HNL}}^Z&=-\frac{\sqrt{g^2 + g'^2}}{2}V_{1\alpha} Z_\mu\left(\overline{\nu}_{1\alpha}^L \gamma^\mu N_1\right)+h.c.,
\end{align}
where $g$ and $g'$ are the weak isospin and weak hypercharge couplings respectively.
Hence, interactions between HNLs and SM fermions are mediated through either charged currents (CC) or neutral currents~(NC). 
For small HNL masses, \ie ${m_{N_1}\ll m_W}$, these interactions can be parametrised through a low-energy effective field theory (EFT) approximation integrating out the heavy mediator particles, $W^{\pm}$ and $Z$.

The HNL also couples to the SM Higgs boson, $H$, through~\cite{Li_2023}
% %
\begin{equation}
    \mathcal{L}_{\text{HNL}}^H = -V_{1\alpha} \frac{m_{N_1}}{v}H\left(\overline{\nu}^L_{1\alpha}N_{1}\right)+ h.c.\,.
    \label{eq:HInteractions}
\end{equation}
However, this contribution tends to be significantly smaller than CC and NC contributions in Equations~\eqref {eq:cc} and~\eqref{eq:nc} for the range of $V_{1\alpha}$ and $m_{N_1}$ considered in this paper; also here an EFT theory approach is used.
%
% Here $g$ is the coupling constant for the $W_{1,2,3}$ bosons under $SU(2)$ and $g'$ for the $B$ boson under $U(1)$.
% All interaction terms are given for a linear dependence in $|V_{1\alpha}|^2$.
%
%

\subsection{Production of Heavy Neutral Leptons}

At hadron colliders, the production of HNLs proceeds through a variety of mechanisms, including decays of hadrons and charged leptons, as well as through electroweak processes such as Drell-Yan production, gluon-gluon fusion, and vector boson fusion involving $W\gamma$ or $Wg$. 
HNL production through hadron decays tends to dominate for $m_{N_1}\ll m_W$ due to the large hadron production cross section at the LHC~\cite{PhysRevD.110.030001}.
%with respect to the cross-sections for lower centre-of-mass energies. 

Hadronic HNL production occurs via semileptonic hadronic decays of the form $h \to h' \ell N_1$,
where $h$ and $h'$ are hadrons and $\ell=e,\mu,\tau$~\cite{Bondarenko:2018ptm}.
Depending on the mass of the HNL, production via different hadrons becomes kinematically accessible, which in turn affects the production branching ratio. 
Additionally, the heavier the initial hadron, the lower the HNL production rate from hadronic decays is, as the hadron's decay width also decreases at larger hadron masses~\cite{Bondarenko:2018ptm,Gorbunov:2007ak}. 
Therefore, the lighter hadrons will typically dominate the production and are copiously produced at the LHC and so could contribute significantly to the sensitivity of the \anubis detector to HNLs. 
However, light hadrons are typically produced in jets, so the event topology can mimic one of the primary backgrounds that arise from jets that are not fully contained within the \atlas detector. Therefore, the selections that will be detailed in Section~\ref{sec:SETacceptance} are optimised to reduce this background by requiring LLP candidates to be separated from significant hadronic activity; as a consequence, HNLs produced in hadronic decays are strongly rejected.
% However, the selections that will be detailed in Section~\ref{sec:SETacceptance} require LLP candidates to be separated from significant hadronic activity, which is found to reject HNLs produced in hadronic decays. 
Removing this selection requirement has a direct bearing on the data-driven background estimate from Ref.~\cite{ANUBIS:2025sgg}, and requires further study.
Therefore, the exploration of \anubis' sensitivity to hadronic production modes alongside production via gluon-gluon fusion is left for future investigation and is not considered further in this article. 
%This approach may be revised in the future by adjusting the selections during data analysis, making use of information from \atlas, or improved understanding of the expected \anubis backgrounds via data or simulations. 
First studies that do not impose any requirement for LLP candidates to be separated from significant hadronic activity suggest that \anubis could achieve excellent sensitivity to HNLs produced in hadronic decay channels~\cite{Hirsch:2020klk,Wang:2025esc}.  

\begin{figure}
    \centering
    \includegraphics[width=0.5\columnwidth]{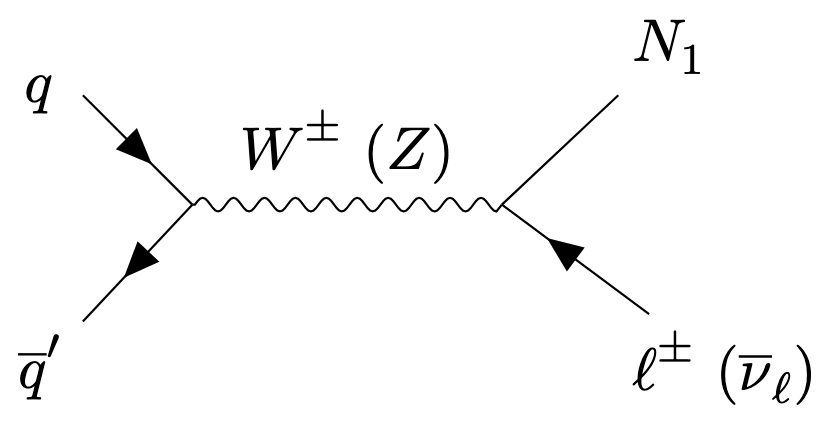}
    \includegraphics[width=0.5\linewidth]{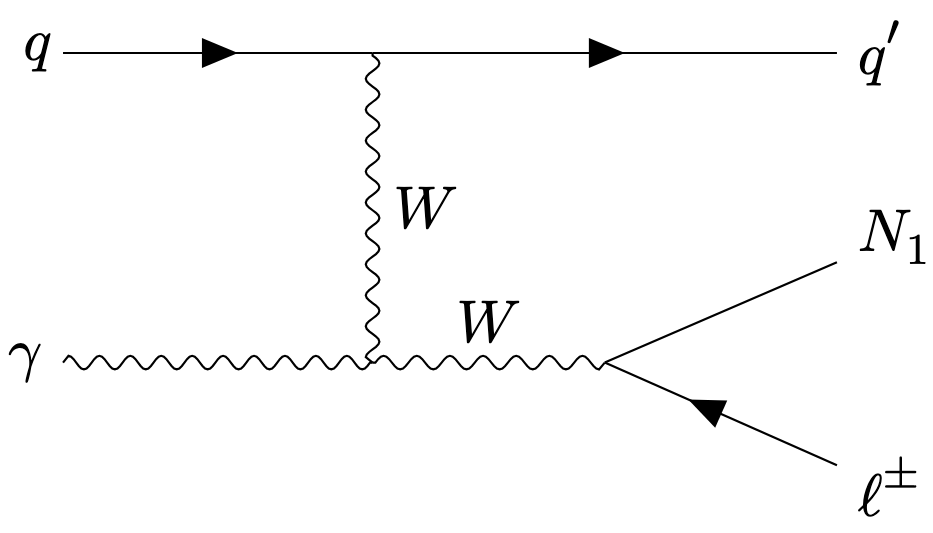}\includegraphics[width=0.5\linewidth]{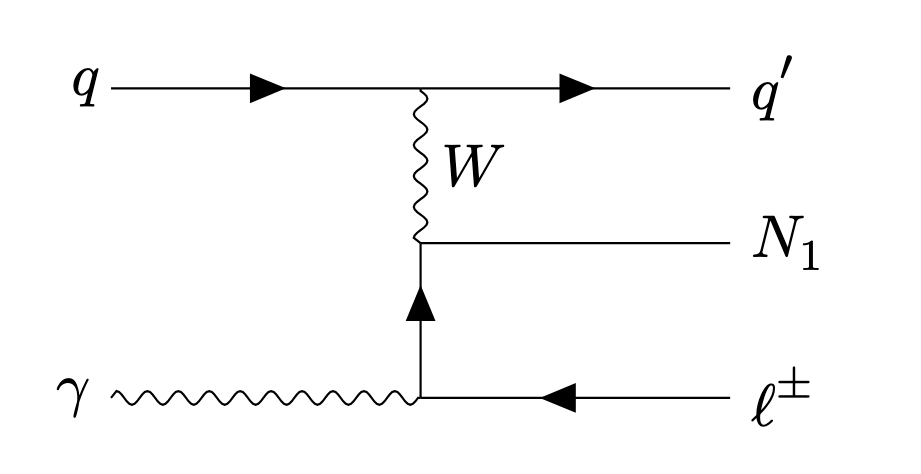}
    \centering
    \includegraphics[width=0.5\linewidth]{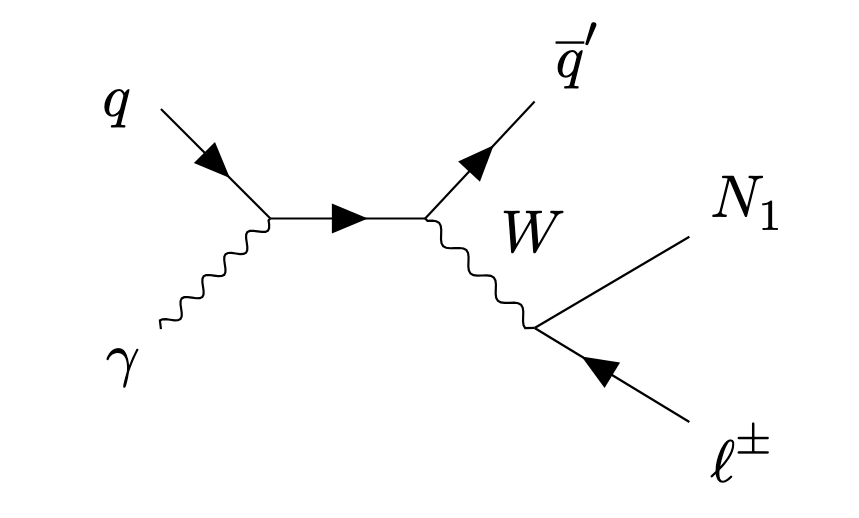}\includegraphics[width=0.5\linewidth]{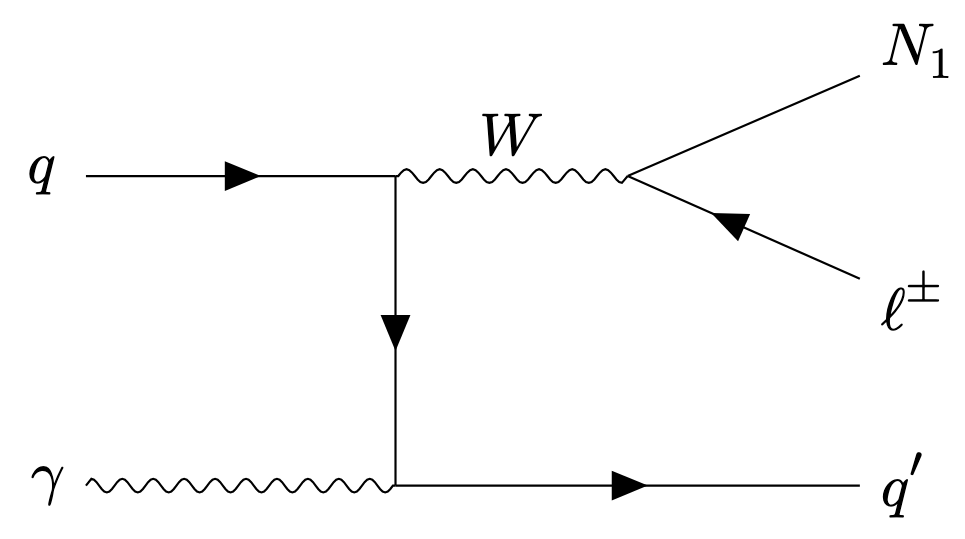}
    \caption{HNL Production modes via off-shell electroweak bosons. The top diagram shows charged and neutral current Drell-Yan production. All other diagrams show $W\gamma$ fusion production processes.}
    \label{fig:WZHproduced_HNL}
\end{figure}

By contrast to HNL production from hadron decays, direct production of HNLs through electroweak gauge bosons $W$ or $Z$ does not produce nearby hadronic activity, and is hence the main focus of this work. 
The relevant direct HNL production modes considered are:
\begin{eqnarray*}
    q \Bar{q}^\prime \to W^{*\pm} \to N_1 \ell^\pm &:& \text{CCDY},\\
    q \Bar{q} \to Z^* \to N_1 \nu_\ell/\Bar{\nu}_\ell &:& \text{NCDY}, \\
    % g g &\to h^*/Z^* \to N_1 \nu_\ell/\Bar{\nu}_\ell \quad\:\:\:\:\: \text{Gluon-Gluon Fusion}, \\
    q \gamma \to q^\prime \ell^\pm N_1 &:& \text{$W\gamma$ fusion},
    % q g &\to W^*/Z^* q' \to N \ell/\nu q' \:\:\: \text{Quark-Gluon Fusion},
\end{eqnarray*}
where CCDY and NCDY stand for charged and neutral current Drell-Yan production, respectively.
The corresponding Born-level Feynman diagrams are shown in Figure~\ref{fig:WZHproduced_HNL}.
The production cross-sections for these three modes considered were calculated in \pythia{8}~\cite{Bierlich:2022pfr} and \madgraph~\cite{Alwall:2014hca}, and compared to values calculated using formulae from Refs.~\cite{Bondarenko:2018ptm,Degrande:2016aje}. 
The cross-section values were found to agree within uncertainties, and so the \madgraph predictions are used throughout this article, unless explicitly indicated otherwise.
Direct production modes are subdominant for $m_{N_1}<5~\GeV$ due to competing production via hadron decays to HNLs. 

\subsection{Decays of Heavy Neutral Leptons}
For $m_{N_1}$ much below the mass of the $W$ boson, HNLs decays proceed via off-shell electroweak gauge bosons into leptons and hadrons.
%At parton level, decays into quarks and leptons are possible, but subsequent hadronisation results in the quarks forming hadrons.
%
The final states mediated by the CCDY process have the generic form 
$$N_1\to \ell_\alpha\,U \overline{D}\,,$$ 
where Greek subscripts indicate the lepton family and two possibilities are present: $U\overline{D} = \{\nu_\beta \ell^+_\beta\}$ or $\{q \bar{q}^\prime\}$ for up- and down-type quarks, $q$ and $q^\prime$, respectively.
For NCDY the final states have the form 
$$N_1 \to \nu f \Bar{f}\,,$$ 
where $f$ are SM fermions.

For light HNLs, two main categories of decay channels are available after hadronisation: 
two-body decays into a meson and a lepton or neutrino, ${N_1 \to \ell^\pm \, h^\mp}$ or ${N_1 \to \nu \, h^0}$, and three-body decays into fermions.
Here, $h^\mp$ and $h^0$ are charged and neutral hadrons, respectively. 
The partial decay widths for each individual decay channel are taken from Ref~\cite{Bondarenko:2018ptm}.
%The combined decay width for these final states are found using the QCD corrections found from studies of $\tau$ decay.

All final states with a charged particle are, in principle, detectable by \anubis, leaving only the three-neutrino final state $N_1 \to \nu \bar\nu \nu$ invisible.
Due to factors like larger branching ratios and the additional showering that is present for quarks but not for leptons, \anubis will have higher sensitivity to some final states.

%%%%%%%%%%%%%%%%%%%%%%%%%%%%%%%%%%%%%%%
\section{Simulation, Acceptance, and Sensitivity} \label{sec:SETANUBIS} 
%%%%%%%%%%%%%%%%%%%%%%%%%%%%%%%%%%%%%%%
To efficiently and flexibly perform sensitivity studies for a wide range of theoretical models, including the minimal HNL model in this article, the Simulation, accEptance and sensiTivity studies framework for \anubis (\setanubis) has been developed. 
The framework provides a common tool to simulate signal models, apply selection criteria that emulate the response of the \anubis detector, and extract sensitivity limits for the relevant model parameters. 
This functionality is realised through a modular architecture that allows for additional models or analysis components to be incorporated while minimally impacting core functions. 
Therefore, the limits produced with the \setanubis framework can be directly compared, since they share the same base assumptions and the tools that implement them in the context of the \anubis detector.

The functionality of \setanubis can be broken down into three main components: 
\begin{enumerate}
\item
simulation of signal samples;
\item
acceptance determination and event selection;
\item
calculation of sensitivity limits.
\end{enumerate}
An overview of the methodology for these sections using the HNL case as an example will be highlighted in the upcoming sections. 
A more detailed account of the SET-ANUBIS framework is given in Ref~\cite{Erner:2026fpj}.

\subsection{Simulating Signal Samples}
\label{sec:SETsim}
Several Monte Carlo (MC) generators can be utilised to generate samples for various signal models using \setanubis, providing flexibility to use the optimal generator for each particular case. 
Currently, \setanubis uses \madgraph (\madversion{3.5.4})~\cite{Alwall:2014hca} and \pythia{8}~\cite{Bierlich:2022pfr}. 
Each model to be examined through \setanubis is implemented separately.
% For example, if a Dark Photon model is to be generated in \pythia, a dedicated run card needs to be produced for that case. 
The framework then automatically generates scripts to produce simulated MC samples in the HEPMC format~\cite{Buckley:2019xhk}. 
The use of the HEPMC common file format allows for the modular structure of the \setanubis framework, since it is implemented in many MC generators, and downstream components can interface with a HEPMC file. 
%The following paragraphs will provide a more detailed overview of the \madgraph and \pythia processes.

For \madgraph, the implementation of a given LLP model is carried out with the FeynRules~\cite{Alloul:2013bka} package and Universal Feynman Output (UFO) model files~\cite{Degrande:2011ua}. 
These UFO model files contain all the necessary information to generate a signal sample, \eg the Feynman rules and the dynamic production and decay widths, which can be used to calculate both the production and decay branching ratios described by the model. 
To verify the production branching ratios and decay widths produced with the HNL UFO, \textsc{SM\_HeavyN\_NLO}~\cite{Ruiz:2020cjx}, the expressions presented in Ref.~\cite{Bondarenko:2018ptm} were additionally evaluated in \mathematica~\cite{Mathematica} and found to be consistent with those produced with the UFO file.
The generation of LLP signal samples in \madgraph is controlled through a run card for a given model to be produced in $pp$ collisions at 14~TeV. 
The decay is handled via \madspin~\cite{Artoisenet:2012st} and \pythia{8} is used for showering. 

%A set of signal samples with a sufficiently high number of events would be required for analysis. Each sample would have differing input parameters that would affect both the production of the LLP and its decay, \eg the mass of the LLP. To handle this a HTCondor batch system~\cite{condor} was utilised, allowing many simulations to be generated in parallel. Each simulation generates a small number of events ($\sim2000$), which can later be combined to accumulate a large number of generated events for different model parameters.

For \pythia{8}, the LLP models are implemented with a configuration file and a set of Python scripts that calculate the production and decay widths, which need to be defined for the particular model. 
The configuration file defining the properties of the LLP model is produced using a factory method pattern to allow for a variety of models to be included while preserving the same general conditions for flexibility and clarity.

At present, \setanubis does not contain a simulation of the detector response, \eg using \geant~\cite{GEANT4:2002zbu}, which is instead accounted for through the signal detection efficiency $\varepsilon_\text{sig}$. 
Following Ref.~\cite{ANUBIS:2025sgg}, $\varepsilon_\text{sig}=0.5$ is taken, where it corresponds to the worst case considered --- displaced vertices with low track multiplicity. 
HNL decays likewise yield low-multiplicity vertices, therefore the same $\varepsilon_\text{sig}$ value is adopted here.
The simulation of detector response is being implemented and is left to future publications.

%This takes a minimal set of inputs for the model, \eg LLP mass and production modes, to populate the file with the added model particles, their production and decay channels. The branching ratios of the latter can be calculated using bespoke Python scripts informed by theory predictions for the model or using tools such as \madgraph or \marty. The samples are then produced by providing the relevant configuration files to \pythia{8}, with the output in HEPMC format.

\subsection{Acceptance Determination and Event Selection}
\label{sec:SETacceptance}
The generated signal samples contain the LLP decay products in the context of a given signal model. 
To determine the sensitivity of \anubis to this model, a set of selections is then applied in \setanubis following the approach established in Ref.~\cite{ANUBIS:2025sgg}.
These selections are briefly summarised in the following paragraphs.

First, the geometric acceptance is determined, which requires the displaced vertex from the LLP decay to be located within the air-filled volume between \atlas and \anubis, as indicated in Figure~\ref{fig:ANUBISAcceptance}, following Ref.~\cite{ANUBIS:2025sgg}. 
In addition, the decay products of the LLP must leave at least two charged tracks within \anubis' tracking stations, and each track must hit both the inner and outer tracking station layers. 
This ensures that only final states that would reach the \anubis detector and produce a detectable displaced vertex are considered. 

%As \anubis is planned to be integrated as an \atlas sub-detector, this allows for triggering decisions based on \anubis signals and properties of the reconstructed \atlas event to be used to reject potential backgrounds. 
A generic feature of a neutral signal LLP for \anubis is that it escapes the \atlas detector and thus the energy carried by that LLP will not be detected by \atlas. This can be expressed through the missing transverse momentum, \met, 
which is defined as the negative vector sum of the momenta of the measured final state particles in the transverse plane\footnote{%
ANUBIS adopts the right-handed coordinate system of ATLAS, where its origin is at the nominal interaction point in the centre of the detector and the $z$-axis points along the beam pipe. The $x$-axis points from the IP to the centre of the LHC ring, and the $y$-axis points upwards.
Polar coordinates ($r$, $\phi$) are used in the transverse plane, $\phi$ being the azimuthal angle around the $z$-axis. 
The pseudorapidity is defined in terms of the polar angle $\theta$ as $\eta \equiv-\ln \tan(\theta/2)$ and is equal to the rapidity $y = \frac12 \ln \left(\frac{E+p_z}{E-p_z}\right)$
in the relativistic limit.
Angular distance is measured in units of $\Delta R \equiv \sqrt{\Delta\eta^2+\Delta\phi^2}$.
}~\cite{ATLAS:2018txj}. Therefore, to select the signal LLP events where there is also some prompt activity within the detector, some \met is required. However, the main sources of background for \anubis are expected to be from hadronic interactions of SM LLPs like $K_L$ and neutrons with material inside the \atlas cavern. These SM LLPs will have a non-zero but typically small \met due to being partially contained within \atlas before they escape. Therefore, to suppress contributions from these low-energy $K_L$ or neutrons and other background sources with small or no missing transverse momentum, $\met>30\,\text{GeV}$ is required~\cite{ANUBIS:2025sgg}. 

Finally, a set of isolation requirements is applied to further reduce contributions from $K_L$ and neutrons, exploiting the fact that they are typically produced in association with prompt (distance from the IP, $d_\text{IP}$, is less than $10\,\text{mm}$) hadronic activity close by in $\Delta R$. 
Specifically, LLP trajectories are required to have an angular separation of $\Delta R_\text{LLP,jet}>0.5$ from any jets with $\pt>15~\GeV$ and of $\Delta R_\text{LLP,trk}>0.5$ from any charged particles with $\pt>5~\GeV$.  
%This is done by defining `charged particle' and `jet' objects, and determining if their momentum lies within a certain $\Delta R=\sqrt{(\Delta\eta)^2+(\Delta\phi)^2}$ separation of the LLP candidate's momentum, $\Delta R=0.5$. 
For these purposes, `jets' are reconstructed using the anti-k$_t$~\cite{Cacciari:2008gp} jet algorithm with a cone parameter of $R=0.4$ from all prompt final-state particles with $|\vec p\,|>0.1~\GeV$ produced within $d_\text{IP}<10\,\text{mm}$, which are not neutrinos, and not associated with the decay of the LLP. 
`Prompt charged particles' are defined as any final-state particle with a non-zero electric charge $q\neq0$ produced close to the primary interaction point $d_\text{IP}<10\,\text{mm}$, and not associated with the LLP decay. 
These definitions are summarised in Table~\ref{tab:PartDefinition}.

\begin{table}[ht]
\caption{Summary of the definition of objects considered in the selections. 
The jets are reconstructed using the anti-k$_t$~\cite{Cacciari:2008gp} jet algorithm with $R=0.4$ from prompt final-state particles that pass the selections outlined below.
}
\label{tab:PartDefinition}
\begin{tabular}{ll}
\hline

\textbf{Objects} & \textbf{Selections} 
\\ 
\hline
\hline
\multirow{2}{*}{Prompt charged particles}\quad\hbox{}     & $q\neq0$, $d_\text{IP}<10~\mm$, \\
 & $|\vec p|>0.1~\GeV$, $\pt>5~\GeV$ \vspace{0.15cm}\\
 % & ID$_\text{PDG}\notin[12,14,16,18]$ \vspace{0.15cm}\\ 
\hline
\multirow{3}{*}{Prompt particles}     & $d_\text{IP}<10~\mm$, \\
& $|\vec p|>0.1~\GeV$, \\
& ID$_\text{PDG}\notin[12,14,16,18]$ \vspace{0.15cm} \\ \hline
Jets                  & $\pt>15~\GeV$                                                       \bigstrut \\ \hline
\end{tabular}
\end{table}

A summary of the selection strategy is shown in Table~\ref{tab:Selection}. 
After applying this selection to the simulated signal samples, the geometric and kinematic acceptance, \acc, can be determined for the LLP signal sample as the ratio of the number of surviving events to the number of events before any selections were applied.

\begin{table*}[ht]
\caption{
Summary of the selection strategy implemented in \setanubis for LLP candidates to determine the geometric and kinematic acceptance. 
The $(x,y,z)$ positions of the decay vertices are determined relative to the ATLAS IP; $r_\text{CoC}$ represents the radius from the centre of curvature of the ATLAS cavern ceiling ($x=-1.7~\metre$, $y=3.52~\metre$); and $r_\text{IP}$ is the radius from the ATLAS IP. 
%($x=1.7~\metre$, $y=-2.38~\metre$, $z=0~\metre$). 
The selections are applied successively in the order shown.}
\label{tab:Selection}
\begin{tabular}{lll}
\hline
\textbf{Category} & \textbf{Description}      & \textbf{Requirement}                
\\ \hline 
\hline 
\multicolumn{1}{l|}{\multirow{5}{*}{\textbf{Geometric acceptance}}} &
  \multirow{2}{*}{LLP decay within the ATLAS cavern} &
  $-17.2<x<13.8~\metre$,  $y > -11.37~\metre$, $|z| < 26.4~\metre$,\\ 
\multicolumn{1}{c|}{} & & $ r_\text{CoC} < 20~\metre$ \\
\multicolumn{1}{c|}{} & ATLAS veto               &  Not within ($|z|<22~\metre$ and $r_\text{IP} < 9.5~\metre$)                  \\
\multicolumn{1}{c|}{} & LLP-ANUBIS intersection       & $N_\text{inters.} \geq 2$ layers                     \\
\multicolumn{1}{c|}{} & LLP Decay charged-track multiplicity              & $N_\text{tracks}\geq2$, $N_\text{hit} \geq 2$ layers per track \\ \hline
\multicolumn{1}{l|}{\multirow{4}{*}{\textbf{Kinematic acceptance}}} &
LLP momentum & $|\vec p_\text{LLP}| > 0.1$~\GeV\\
\multicolumn{1}{c|}{} & \met requirement               & $\met > 30~\GeV$                                \\
\multicolumn{1}{c|}{} & LLP isolation from jets              & $\Delta R_\text{LLP,jet} > 0.5 $                           \\
\multicolumn{1}{c|}{} & LLP isolation from energetic tracks & $\Delta R_\text{LLP,trk} > 0.5$                        \\ \hline
\end{tabular}
\end{table*}

\subsection{Determining Sensitivity Limits}
\label{sec:SETsensitivity}
The last stage of \setanubis is to obtain the projected sensitivity estimates. 
Following the conventions of the PBC initiative,~\cite{PBC:2025sny}, the exclusion sensitivity using the $CL_\text{s}$ method~\cite{Read:2002hq} at 95\% CL is used for benchmarking the ANUBIS sensitivity to potential BSM LLP models.
To this end, ANUBIS is treated as a counting experiment, and a sensitivity threshold of 4 candidate events~\cite{PhysRevD.110.030001} is considered, 
%$N_\text{LLP}\geq4$
 assuming that the HNL search using ANUBIS is background-free.
This is a reasonable assumption since HNL production at the LHC is always accompanied by a prompt lepton, as shown in Figure~\ref{fig:WZHproduced_HNL}. 
The presence of a prompt lepton allows a strong reduction of backgrounds from QCD multijet production and hence reduces the expected number of background events of $\mathcal{O}(100)$ in a conservative background estimation from Ref.~\cite{ANUBIS:2025sgg}, by several orders of magnitude.
%is required to determine a 95\% CL exclusion sensitivity. 
%This background estimate was based on an \atlas LLP analysis~\cite{ATLAS:2018tup}, however a small-scale demonstrator unit, \proanubis~\cite{ANUBIS:2025mme,ANUBIS:2025inn}, has been taking data in the \atlas cavern since early 2024 and so is directly measuring the expected background level. In future, a data-driven estimate of the background could be used as the moderate background threshold. Alternatively, a detailed simulation study of the backgrounds in a \geant~\cite{GEANT4:2002zbu} model that includes both ATLAS and the ANUBIS detector could also provide an alternative background estimate. 
In \setanubis, the required sensitivity thresholds on the number of expected events from LLP decays, $N_{LLP}$, can be assigned directly by the user.

The value of $N_{LLP}$ is determined by
\begin{align}
    N_\text{LLP} = \mathcal{L}\,\sigma_\text{LLP}\,\BF_\text{LLP}\,\acc\,\varepsilon_\text{sig}\,,
    \label{eq:NLLP}
\end{align}
where $\mathcal{L}$ is the integrated luminosity, $\sigma_\text{LLP}$ is the production cross-section of the LLP, and $\BF_\text{LLP}$ is the branching ratio of the LLP into the considered final state(s). 
%, and $\varepsilon_{sig}$ is the efficiency with which \anubis could observe a signal event. 
In this study, $\mathcal{L} = 3\,\text{ab}^{-1}$ is assumed, in alignment with the conventions in the transverse and forward detector communities as motivated by the expectations at the High-Luminosity LHC (HL-LHC).
%and $\varepsilon_\text{sig} \equiv 50\%$ is assumed, following Ref.~\cite{ANUBIS:2025sgg} where this is a conservative value taken when deriving the projected background level.
The sensitivity limits can then be cast into different parameters of interest.
Commonly this is done either by directly setting $N_\text{LLP}$ to the sensitivity threshold, \eg 4, and calculating the associated $\sigma_\text{LLP}$, $\BF_\text{LLP}$ or $\sigma_\text{LLP}\BF_\text{LLP}$ value for a particular set of model parameters, \eg $m_{N_1}$ and $|V_{1\alpha}|^2$. 
Alternatively, $N_\text{LLP}$ can be calculated directly for each subset of $N$ model parameters, to produce an $N$-dimensional histogram where each bin is assigned the associated $N_\text{LLP}$. 
Then the sensitivity limits are defined by the contour where $N_\text{LLP}$ is equal to the sensitivity threshold.

The modular nature of \setanubis ensures that Equation~\eqref{eq:NLLP} can be adapted for each model. 
However, the baseline of the sensitivity threshold should remain invariant to allow for easier comparison of limits from different LLP models.

%%%%%%%%%%%%%%%%%%%%%%%%%%%%%%%%%%%%%%%
\section{Event Generation} \label{sec:EventGen} 
%%%%%%%%%%%%%%%%%%%%%%%%%%%%%%%%%%%%%%%
This study considers the addition of a single Majorana HNL that couples solely to either the electron (${|V_{1e}|:|V_{1\mu}|:|V_{1\tau}|=1:0:0}$) or muon (${|V_{1e}|:|V_{1\mu}|:|V_{1\tau}|=0:1:0}$). 
These choices correspond to BC 6 and BC 7 scenarios from the PBC initiative~\cite{Beacham:2019nyx,Alemany:2019vsk}, respectively; the solely tau coupled BC~8 scenario was reserved for future study. 
Signal HNL events were generated using the \textsc{SM\_HeavyN\_NLO}~\cite{Ruiz:2020cjx} UFO file in \madversion{3.5.4} with the NN23LO1 PDF set~\cite{Ball:2013hta}, through three production modes: CCDY, NCDY, and W$\gamma$ fusion ($W\gamma$). 
The produced $N_1$ was then forced to decay with ${\BF(N_1\rightarrow X)=100\%}$ into either $X=q\bar{q}'\ell$, $X=q\bar{q}\nu_{\ell}$, $X=\ell\bar{k}\nu_k$, or $X=k\bar{k}\nu_{\ell}$, where $\ell=e$ or $\mu$, ${k=e,\mu}$ or $\tau$, and the prime indicates weak-isospin conjugate-states, using Madspin. 
The subsequent hadronisation and parton shower were simulated with \pythia{8.311}. 
This leads to a set of nine signal categories corresponding to the combination of each production and decay modes, \eg CCDY production with $N_1\rightarrow q\bar{q}\nu_{\ell}$. 

The two model parameters considered in this study are free parameters of the HNL model: $m_{N_1}$ and $|V_{1\ell}|$. 
The HNL coupling parameter was scanned in the range of
%at set values for each of nine signal categories: 
%$|V_{1\ell}|\in$[1, 0.1, 0.3162, 0.01,  $3.162\times10^{-2}$, $10^{-3}$, $3.162\times10^{-3}$, $10^{-4}$, $3.162\times10^{-4}$, $10^{-5}$, $3.162\times10^{-5}$, $10^{-6}$, $3.162\times10^{-6}$]
${|V_{1\ell}|\in[1, 10^{-6}]}$ in 12 logarithmically equidistant steps, supplemented by a more granular grid of ${|V_{1\ell}|\in[1.41,1.732,2,2.449,2.646,2.828,3]\times10^{-4}}$ corresponding to the region of maximum sensitivity. 
%these values were chosen for their coverage in $|V_{1\ell}|^2$; 
The HNL mass parameter was studied in the ranges of  ${m_{N_1}\in[0.2, 1.1]}$~\GeV in steps of 0.1~\GeV, ${m_{N_1}\in[1.3, 3.7]}$~\GeV and ${m_{N_1}\in[5.7, 9.1]}$~\GeV in steps of 0.2~\GeV, and ${m_{N_1}\in[4.1, 5.3]}$~\GeV in steps of 0.4~\GeV. 
For each set of these ($|V_{1\ell}|$, $m_{N_1}$) parameters, $\mathcal{O}(10^4)$ events were generated for both the electron and muon coupled models.
% 6000 (16000) for unreweighted cases and 36000 (96000) for reweighted cases for the electron (muon) coupled models.

\begin{figure}[ht]
    \centering
    \includegraphics[width=0.95\linewidth]{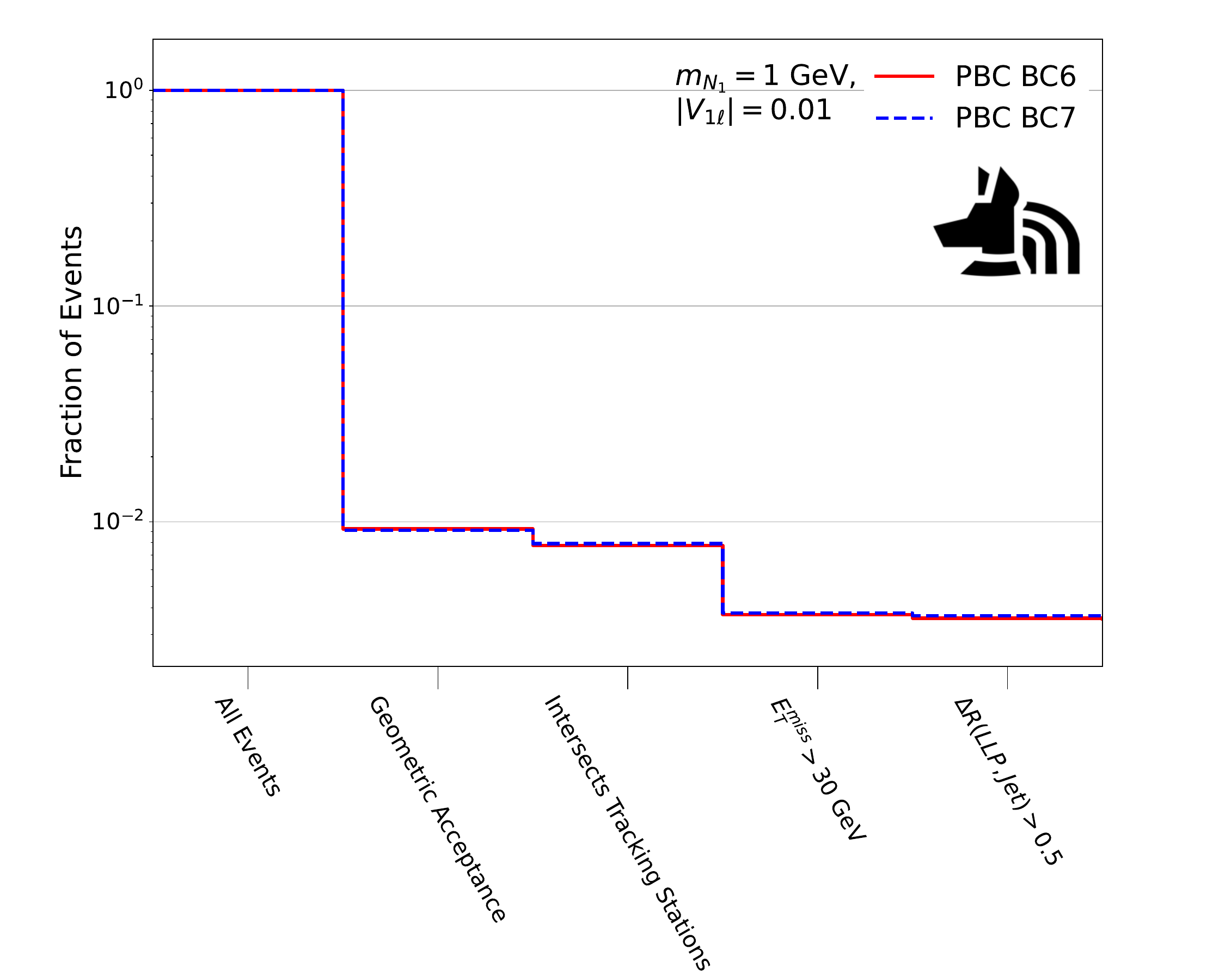}
    \caption{Fraction of surviving signal events 
    %summed by the production and decay modes after 
    applying the successive selections shown in Table~\ref{tab:Selection} for BC6 and BC7, where $m_{N_1}=1~\GeV$ and $|V_{1\ell}|=0.01$.}
    \label{fig:CutFlows}
\end{figure}

To identify the events that would be consistent with an HNL decay within the \anubis detector, the selection requirements from Table~\ref{tab:Selection} were applied. 
Their impact on the signal dataset for $m_{N_1}=1~\GeV$ and $|V_{1\ell}|=0.01$ is shown in Figure~\ref{fig:CutFlows} as a representative example.
Overall, the selection has an almost identical effect on both BC6 and BC7, with the geometric acceptance being the most impactful part, resulting in $\mathcal{A}\approx4\times10^{-3}$.

%The sensitivity was then evaluated at a threshold of $N_{LLP}=4$, effectively zero background. This is because the HNL is produced in association with a prompt lepton, which would provide a clear trigger within the \atlas detector. The probability of a coincident signal in \atlas and \anubis with the appropriate topology is $\sim10^{-6}$~\cite{ANUBIS:2025sgg}, and therefore can be considered as a low background scenario.

The sensitivity contour is defined by the threshold of $N_{LLP}=4$, corresponding to an effectively background-free search. 
This follows from the HNL being produced in association with a prompt lepton that provides a clear trigger signature within the ATLAS detector. 
The requirement of a prompt lepton reduces the conservative background estimate corresponding to 36 events for exclusion sensitivity from Ref.~\cite{ANUBIS:2025sgg} by a factor $\order{10^{6}}$, \ie the ratio of the total production cross section~\cite{TOTEM:2017asr} to the $W^\pm$ and $Z$ boson production cross section~\cite{ATLAS:2016fij}.

%%%%%%%%%%%%%%%%%%%%%%%%%%%%%%%%%%%%%%%
\section{Results} \label{sec:sens}
%%%%%%%%%%%%%%%%%%%%%%%%%%%%%%%%%%%%%%%

% The exclusion sensitivity of \anubis to the minimal Majorana HNL models, BC 6 and 7, was determined using the method outlined in Section~\ref{sec:SETsensitivity}. 
% In this case, Equation~\eqref{eq:NLLP} is evaluated for each signal category, and for each value of $m_{N_1}$ and $|V_{1\ell}|$, to obtain a contour at $N_{LLP}=4$ in the $m_{N_1}$ vs $|V_{1\ell}|$ parameter-space. The value of $\sigma_{LLP}\BF_{LLP}$ is taken from the output of the \madgraph event generation. 

% The `background-free' sensitivity threshold was selected due to the underlying topology of the HNL production modes (Figure~\ref{fig:WZHproduced_HNL}). In each of the considered cases, the HNL is produced in association with a lepton that could give a clear prompt signal within \atlas. The probability of observing a prompt signal in \atlas that can be associated with a significantly displaced signal in \anubis from background sources is $\sim\mathcal{O}(10^{-24})$~\cite{ANUBIS:2025sgg}. Therefore, the limits shown will be assumed to be background-free.

\begin{figure}[ht]
    \centering
    \includegraphics[width=0.95\linewidth]{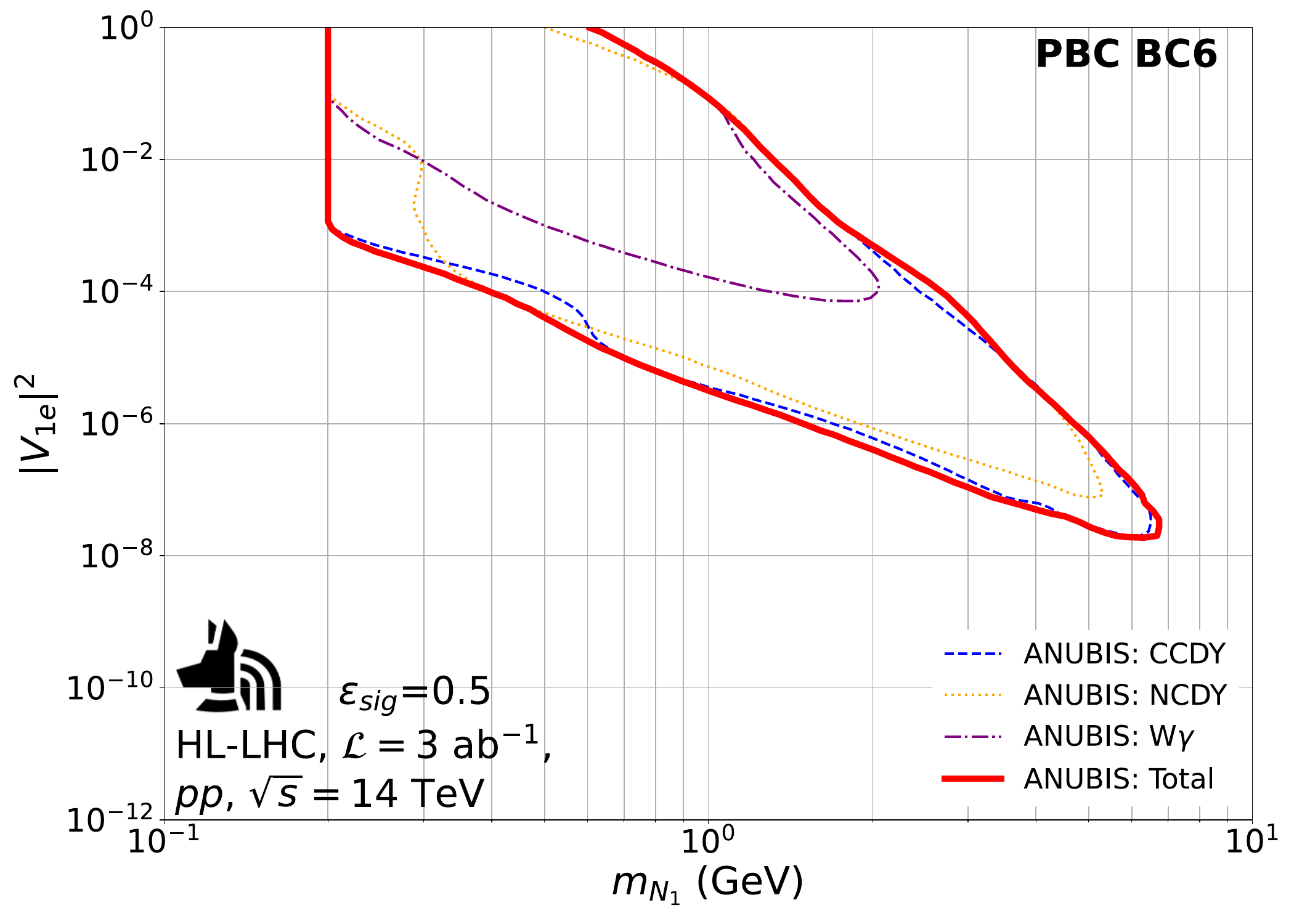}
    \includegraphics[width=0.95\linewidth]{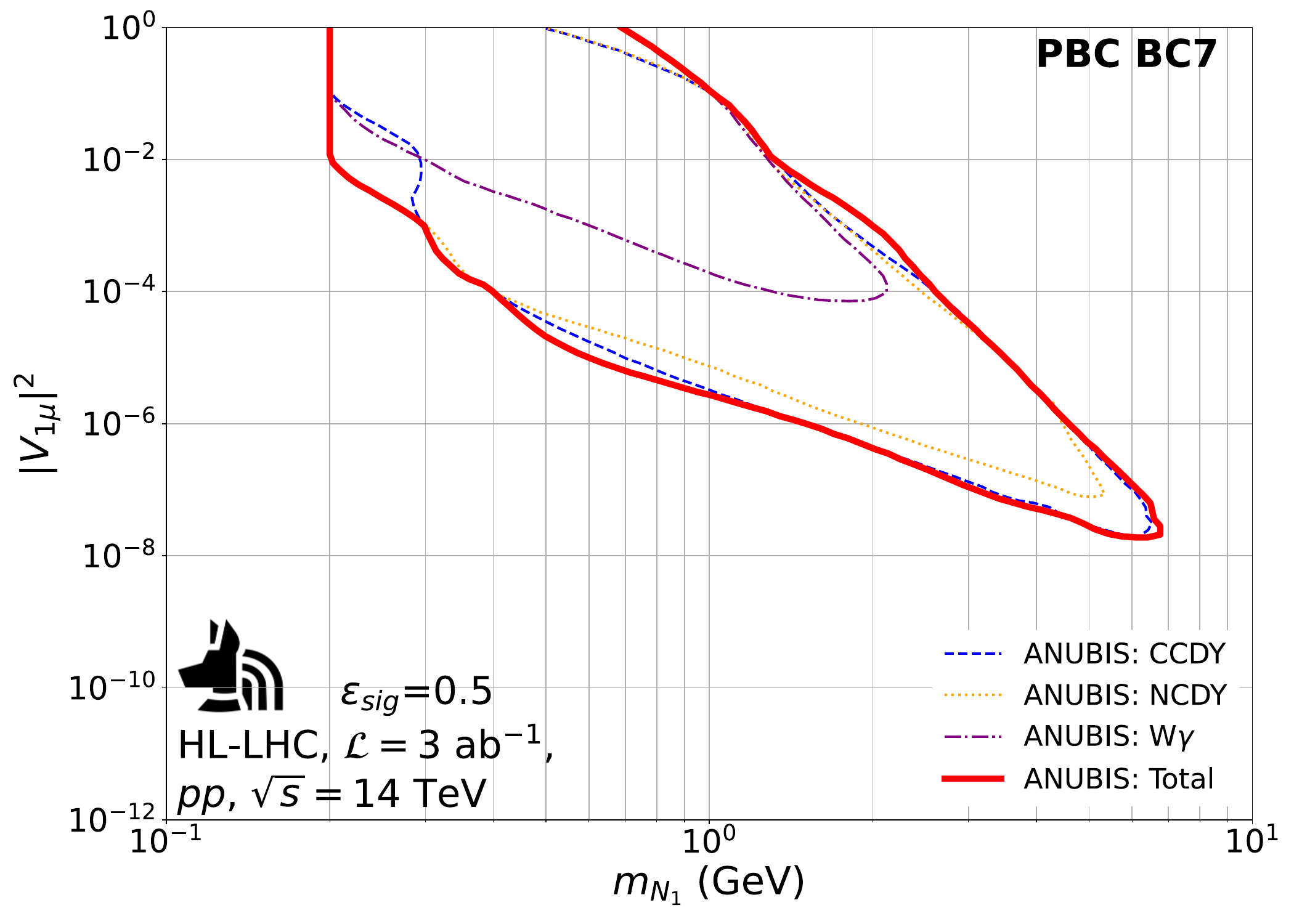}
    \caption{Projected exclusion sensitivity of the \anubis detector to minimal Majorana HNL scenarios with exclusive $|V_{1e}|$ (top) and $|V_{1\mu}|$ (bottom) couplings, corresponding to the BC6 and BC7 PBC scenarios. 
    The combined sensitivity (solid line) is overlaid with the individual contribution from the different production modes (dashed lines).
    }
    \label{fig:HNL_Limits_Combination}
\end{figure}

Figure~\ref{fig:HNL_Limits_Combination} shows the exclusion sensitivity limits for the exclusively electron- and muon-coupled Majorana HNL scenarios, corresponding to the BC 6 and BC 7 PBC benchmarks, respectively.
The limits are shown for a combination of all production and decay modes, overlaid with a breakdown into contributions from the individual decay modes. 
The sensitivity is very similar for electron- and muon-coupled HNL scenarios due to similar acceptances. 
In both cases, the dominant contribution arises from the CCDY production, which is driven by the larger $W$ production cross-section compared to other production modes.
The exclusion sensitivity contour has a fin-like shape in the $\left(\ln\left[|V_{1\ell}| ^2\right],\ln[m_{N_1}]\right)$ space oriented such that the smallest couplings are probed for the largest $m_{N_1}$. 
This dependence is driven by the geometrical acceptance of the ANUBIS detector, which can be understood given that $|V_{1\ell}|^2$ is anti-correlated with the proper decay time $c\tau$, while decreasing $m_{N_1}$ corresponds to a larger relativistic boost factor and hence a larger decay length $\beta\gamma c\tau$. Additionally, at $m_{N_1}=0.2$ BC6 can probe a smaller coupling than BC7, down to $|V_{1e}|^2=10^{-3}$. This is due to kinematic constraints around the di-muon threshold for the muon-coupled scenario. 

%the three production modes had almost identical values for $\mathcal{A}$, so differences can only arise from the $\sigma_{LLP}\BF_{LLP}$ component. 
%

\begin{figure}[ht]
    \centering
    \includegraphics[width=0.95\linewidth]{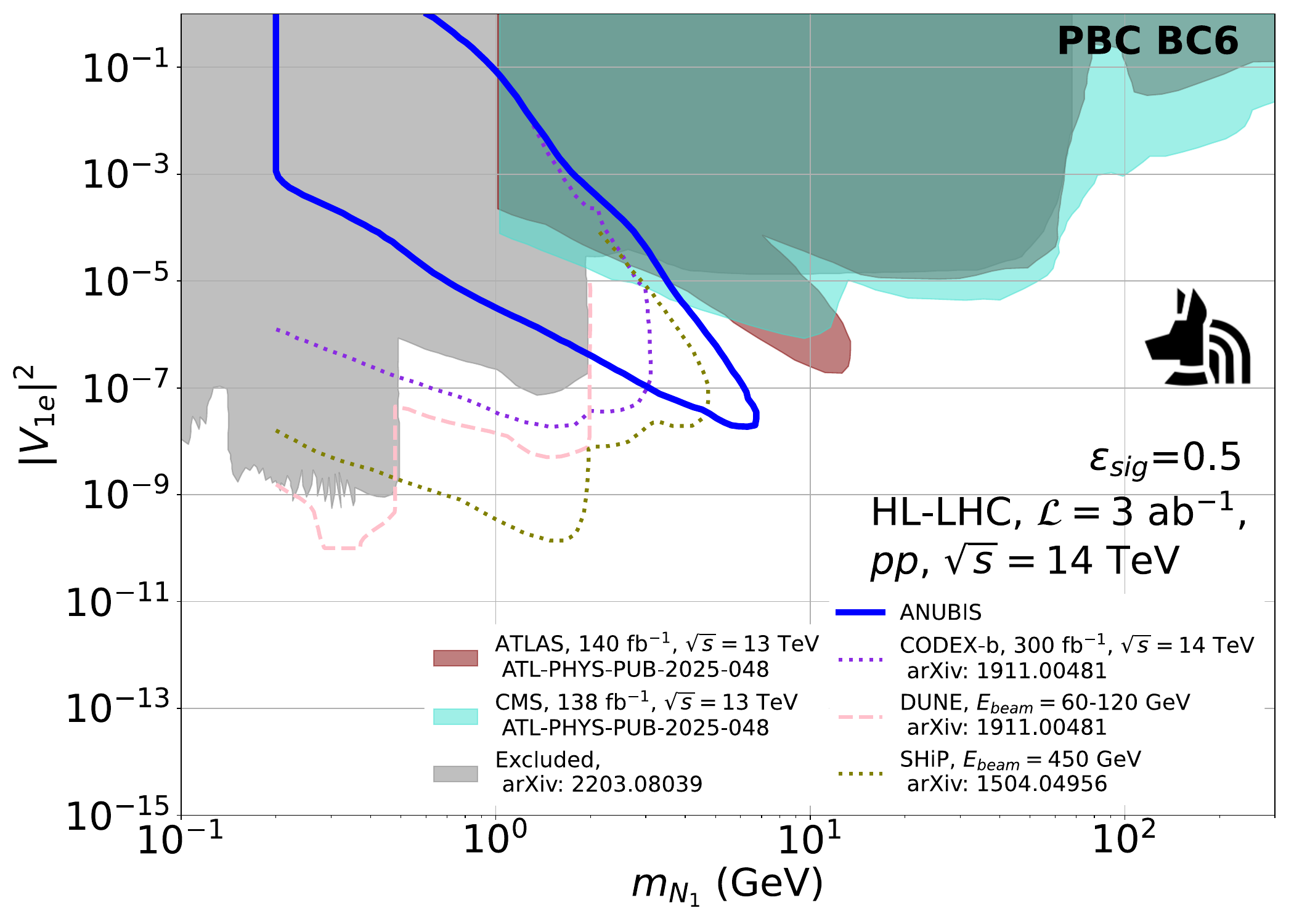}
    \includegraphics[width=0.95\linewidth]{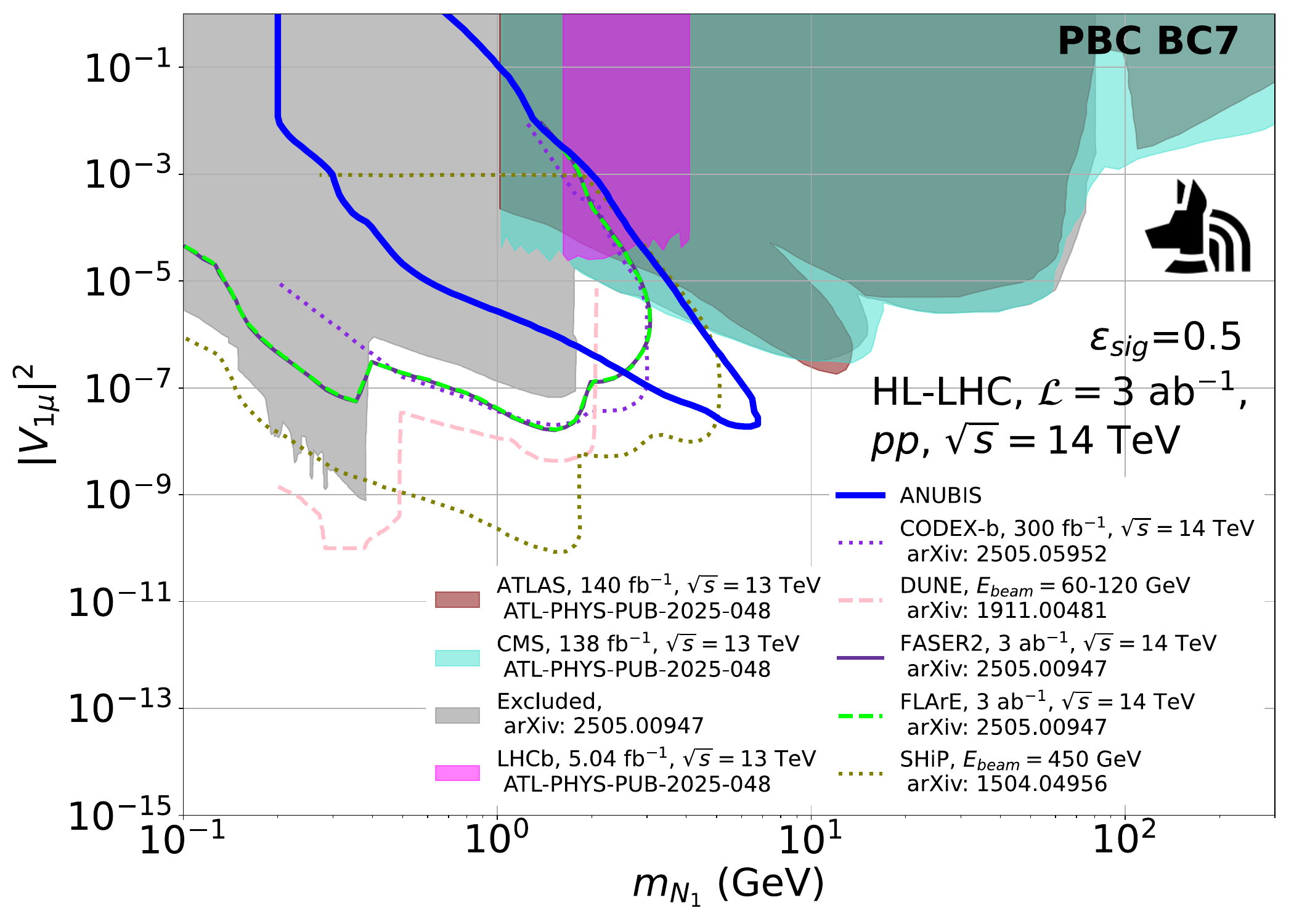}
    \caption{Projected exclusion sensitivity of the \anubis detector to minimal Majorana HNL scenarios with exclusive $|V_{1e}|$~(top) and $|V_{1\mu}|$~(bottom) couplings, corresponding to the BC6 and BC7 PBC scenarios.
    Also shown are projected sensitivities and excluded regions from previous searches summarised in Refs.~\cite{PBC:2025sny,Aielli:2019ivi,ATL-PHYS-PUB-2025-048}.
    }
    \label{fig:HNL_Limits_Overview}
\end{figure}

Figure~\ref{fig:HNL_Limits_Overview} compares the projected ANUBIS exclusion limits to the limits from previous searches and future projections. 
One can observe that \anubis would provide complementary sensitivity to existing constraints on the parameter-space of the model, in particular for $|V_{1\ell}|^2\approx10^{-7}$ and $m_{N_1}\approx5~\GeV$.
Some of this area can be potentially covered with \codex~\cite{Aielli:2019ivi} and SHiP~\cite{SHiP:2021nfo}. 
However, \anubis is expected to provide unique sensitivity at somewhat larger $m_{N_1}\gtrsim5~\GeV$ values, achieving a maximum sensitivity of ${|V_{1\ell}|^2=1.8\times10^{-8}}$ ($1.9\times10^{-8}$) 
at $m_{N_1}=6.4~\GeV$ (6.3~\GeV) for the BC6 (BC7) benchmark scenario. 
Hence, ANUBIS' sensitivity is complementary to SHiP and FASER2, which dominate for $m_{N_1}<1.6~\GeV$ by harnessing HNL production from semileptonic hadron decays.
With the current selection outlined in Table~\ref{tab:Selection} that is dictated by ANUBIS' data-driven background estimate informed by Ref.~\cite{ATLAS:2018tup}, these hadronic HNL production modes are eliminated by the isolation requirement. 
It is expected that a significant fraction of the sensitivity to hadronic HNL production modes can be recovered in data analysis through refined analysis selections and multivariate machine learning techniques, such as utilising detailed \atlas event information to distinguish displaced decays in \anubis from the primary background of material interactions induced by escaped SM LLPs. This would unlock ANUBIS' sensitivity to $m_{N_1}<3~\GeV$. 
A comparable outlook is indicated by the recent study of Wang and Zhang~\cite{Wang:2025esc} that focuses on hadronic production modes of HNLs and compares a number of different experiments including ANUBIS, and which reports a substantial improvement of ANUBIS' sensitivity for $m_{N_1}\lesssim3~\GeV$. 
That study does not include a background treatment, which becomes the limiting consideration once the requirement of isolation from hadronic activity is relaxed; its projections should therefore be regarded as indicative of the potential gain rather than as a quantitative estimate of it.
% This positive outlook is supported by recent studies by Wang and Zhang~\cite{Wang:2025esc}, who focused on hadronic production modes of HNLs and compared them for a number of different experiments including ANUBIS. 
% Their studies showed a significant improvement of ANUBIS' sensitivity for $m_{N_1}\lesssim3~
% \GeV$, leading to ANUBIS outperforming \codex by about an order of magnitude between $m_N=0.1$ and 3 GeV. %Further highlighting that accessing these hadronic modes could also improve sensitivity in the lower mass region.
The refinement of selections to hadronic HNL production is left for future study.

%%%%%%%%%%%%%%%%%%%%%%%%%%%%%%%%%%%%%%%
\section{Conclusions} \label{sec:conc}
%%%%%%%%%%%%%%%%%%%%%%%%%%%%%%%%%%%%%%%
To fully exploit the physics potential of the HL-LHC, a transverse experiment should be installed to capture unique areas of LLP parameter-space characterised by LLPs with decay lengths of $\gtrsim$10~m produced at partonic centre-of-mass energies at or above the electroweak scale.
%in particular for long lifetimes above the electroweak scale, and provide complementarity to other searches. 
The \anubis experiment is a proposed transverse detector that aims to fill such a role. 
The \setanubis framework was developed as a modular and flexible set of software tools to evaluate ANUBIS' sensitivity to a variety of LLP models. 
The framework facilitates the generation of MC events for LLP models, the application of selection criteria that replicate the expected data analysis strategy of \anubis; and the extraction of projected sensitivity limits. 

\anubis' sensitivity to the BC6 and BC7 benchmark scenarios proposed by the PBC working group~\cite{Alemany:2019vsk} was evaluated using the \setanubis framework. 
In particular, a minimal Majorana HNL model is considered where a single new right-handed neutral lepton that couples exclusively to SM left-handed electrons or muons is introduced, corresponding to the BC6 and BC7 scenarios, respectively. 
Overall, \anubis is expected to extend the sensitivity of existing searches in the region of ${3\lesssim m_{N_1}\lesssim 8~\GeV}$ and ${10^{-8}\lesssim|V_{1\ell}|^2\lesssim10^{-5}}$.
Moreover, \anubis is shown to have unique reach among any proposed experiments for ${m_{N_1}\approx5~\GeV}$ and ${|V_{1\ell}|^2\approx10^{-8}}$, reaching its maximum sensitivity of ${|V_{1\ell}|^2=1.8\times10^{-8}}$ ($1.9\times10^{-8})$ at ${m_{N_1}=6.4~\GeV}$ (6.3~\GeV) for BC6 (BC7).

%Other studies~\cite{Hirsch:2020klk,Wang:2025esc} imply that HNLs produced from the decays of B and D mesons should provide better sensitivity. However, these cases were eliminated through the current analysis strategy. It is expected that \anubis could dramatically increase its sensitivity to HNLs, both to lower couplings and to higher HNL masses, by accessing these hadronic decay production modes, which could be achieved through the introduction of machine learning tools and refinement of the selection strategy during actual data-taking.
%%%%%%%%%%%%%%%%%%%%%%%%%%%%%%%%%
%%%%%%%%%%%%%%%%%%%%%%%%%%%%%%%%%
\section{Acknowledgements} \label{sec:ack}
%%%%%%%%%%%%%%%%%%%%%%%%%%%%%%%%%
%%%%%%%%%%%%%%%%%%%%%%%%%%%%%%%%%
We thank Benjamin Fuks for his advice when considering the HNL models and their MadGraph implementations. Additionally, thanks to IPPP for hosting the SET-ANUBIS framework on its GitLab service during development. Also, we gratefully acknowledge the support of UKRI under the Future Leaders Fellowship scheme.

\clearpage

\bibliographystyle{JHEP}
\bibliography{bibs/theory,bibs/general,bibs/anubis-standard-refs}
\clearpage
\onecolumngrid
\appendix
% \subfile{Appendices/HNL_Decays}
\section{Auxiliary Material}
\label{app:AuxiliaryMaterial}
%Testing to see if adding a long stretch of text to this section helps resolve the column issue and get the figure to be properly located on the same page as the section title itself.
\begin{figure*}[htb]
    \centering
    \includegraphics[width=0.48\linewidth]{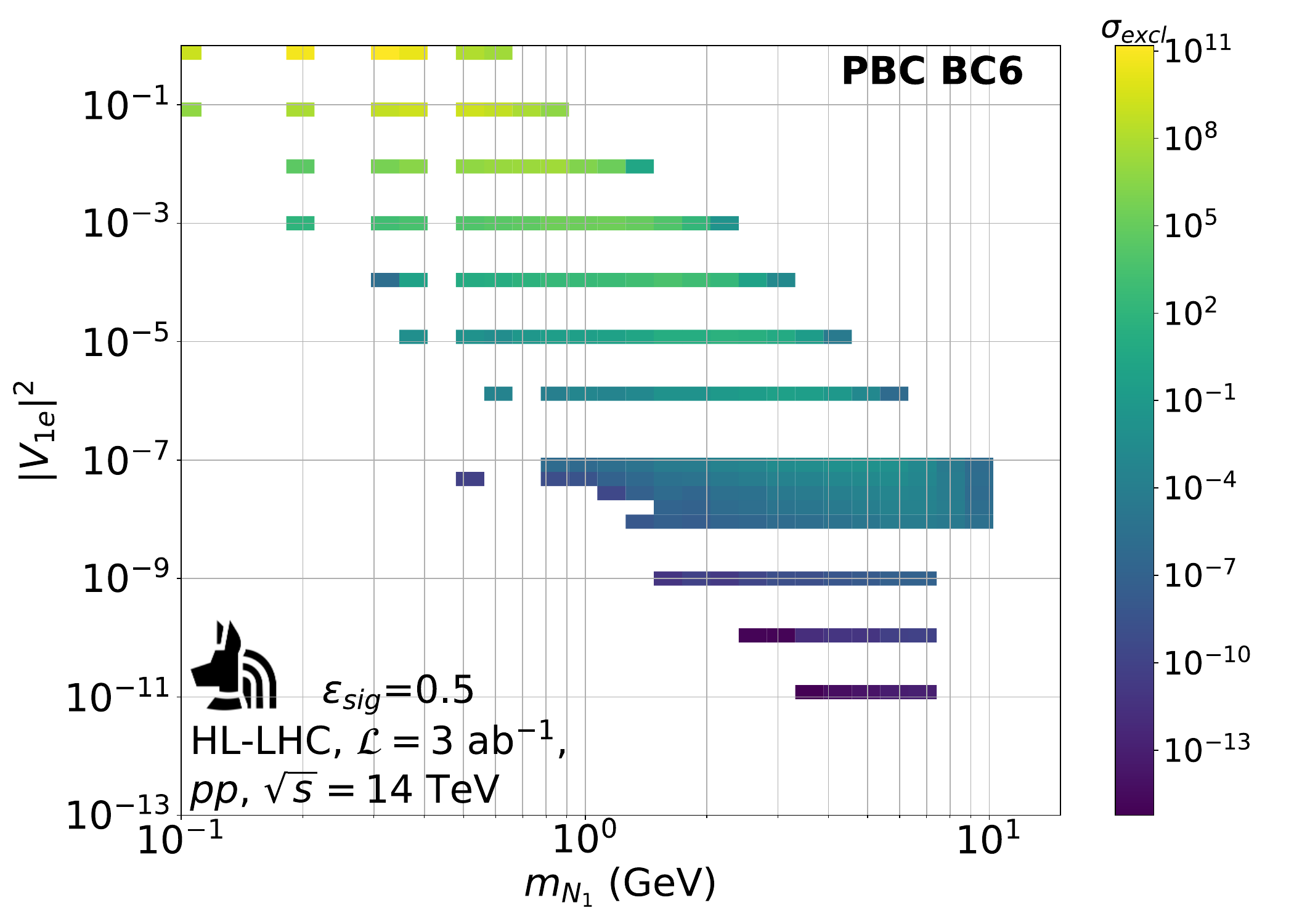}
    \includegraphics[width=0.48\linewidth]{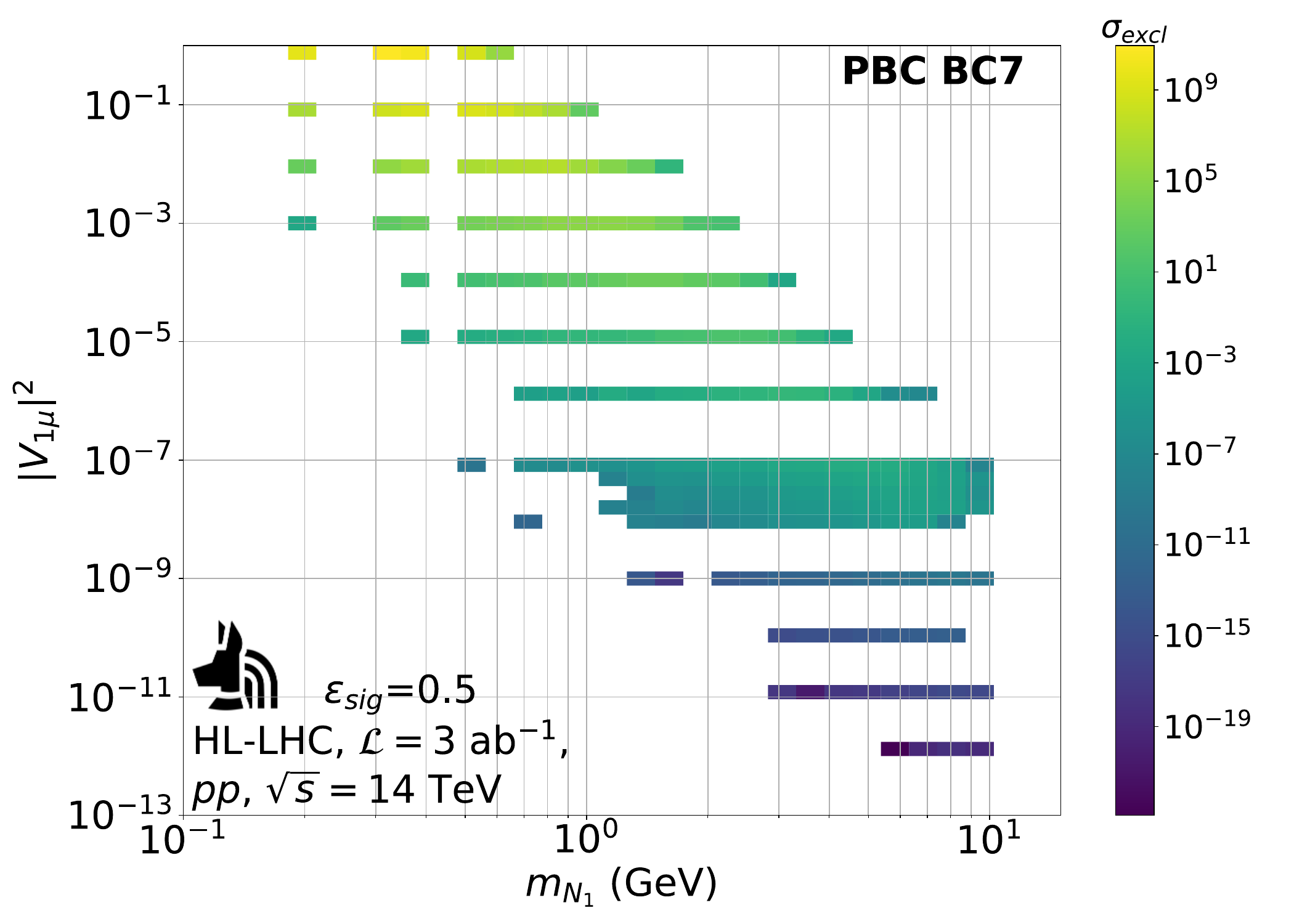}  
    \caption{Projected excluded production cross-sections for minimal Majorana HNL scenarios with exclusive $|V_{1e}|$~(left) and $|V_{1\mu}|$~(right) couplings, corresponding to the BC6 and BC7 PBC scenarios for ANUBIS.}
    \label{fig:ExcludedCross-sections}
\end{figure*}

\end{document}